\documentclass[twocolumn, aps, prl]{revtex4-1}
\usepackage{amsmath,graphicx,float,mathrsfs}

\begin{document}

\title{Universality of Energy Equipartition in One-dimensional Lattices}

\author{Weicheng Fu}
\author{Yong Zhang}
\author{Hong Zhao}
\email{zhaoh@xmu.edu.cn}

\affiliation{Department of Physics and Jiujiang Research Institute, Xiamen University, Xiamen 361005, Fujian, China}

\date{\today }
\begin{abstract}
We show that a general one-dimensional (1D) lattice with nonlinear inter-particle interactions can always be thermalized for arbitrarily small nonlinearity in the thermodynamic limit, thus proving equipartition hypothesis in statistical physics for an important class of systems. Particularly, we find that in the lattices of interaction potential $V(x)=x^2/2+\lambda x^n/n$ with $n\geq4$, there is \textit{a universal scaling law} for the thermalization time $T^{eq}$, i.e., $T^{eq}\propto\lambda^{-2}\epsilon^{-(n-2)}$, where $\epsilon$ is the energy density. Numerical simulations confirm that it is accurate for an even $n$. A slight correction is needed for an odd $n$, which is due to the Chirikov overlap occurring in the weakly nonlinear regime between extra vibration modes excited by the asymmetry of potential. Based on this scaling law, as well as previous prediction for the case of $n=3$, a universal formula for the thermalization time for a 1D lattice with a general interaction potential is obtained.

\end{abstract}

\maketitle

{\it{Introduction}}.---Equipartition hypothesis assumes that an arbitrary small nonlinearity is enough to thermalize a macroscopic thermodynamic system, i.e., the energy will be equally distributed among the various Fourier modes. It is the foundation of statistical physics. The pioneering numerical experiments by Fermi, Pasta, Ulam (FPU), and Tsingou \cite{osti_4376203,dauxois:ensl-00202296}, showed very little tendency toward equipartition of energy among the degrees of freedom, known as the FPU paradox \cite{osti_4376203}. Their seminal work has stimulated a huge amount of research (see Refs. \cite{2008LNP728G,PhysRevLett.98.047202,PhysRevX.4.011054,PhysRevE.93.022216,PhysRevLett.117.163901,PhysRevE.95.060202,PhysRevX.7.011025,NaturePhotonics.12.303,PhysRevE.97.012143,doi:10.1063/1.5009492,PhysRevX.8.041017} and references therein). Extensive numerical simulations have shown clear evidence that there is a energy threshold above which the FPU system reaches a fast thermalized state \cite{PhysRevA.31.1039,FORD1992271,PhysRevE.55.6566,PhysRevE.60.3781}. However, whether a system can be generally thermalized for arbitrary small nonlinearity has not been settled clearly due to the difficulty of rigorous mathematical proof \cite{2008LNP728G}. Recently, resonant wave-wave interaction theory \cite{1992kstbookZ,Majda1997,ZAKHAROV2001573,ZAKHAROV20041,2011LNP825N,2018htdbookZ} has been applied to this problem \cite{Onorato4208,PhysRevLett.120.144301,0295-5075-121-4-44003}.

Wave resonance theory, developed from the statistical mechanics theory of wave turbulence \cite{1992kstbookZ,Majda1997,ZAKHAROV2001573,ZAKHAROV20041,2011LNP825N,2018htdbookZ}, provides a framework for drawing a firm conclusion. This theory assumes that, in the weakly nonlinear regime, the long time dynamics is determined by exact resonances. Particularly, the irreversible transfer of energy resulting in thermalization is achieved by the nontrivial resonance of the Umklapp process. To characterize the nontrivial resonance, one should rewrite the equation of motion with proper canonical variables of normal modes in the Fourier space. If a nontrivial resonance can be found, then the equipartition time $T^{eq}$ is estimated using the amplitude of the resonance. Otherwise, higher harmonics of the normal modes are considered to find the higher order resonance, and then the equipartition time is determined by the amplitude of the higher order resonance \cite{Onorato4208}. Landmark progress has been made recently in lattice models with Hamiltonian
\begin{equation}\label{eqH}
    H= \sum_j \frac{p_j^2}{2}+\frac{(q_{j+1}-q_j)^2}{2}+\frac{\lambda}{n} (q_{j+1}-q_j)^n
\end{equation}
for the special cases of $n=3$ (FPU-$\alpha$ model) \cite{Onorato4208} and $n=4$ (FPU-$\beta$ model) \cite{PhysRevLett.120.144301}, where $p_j$ and $q_j$ denote the momentum and the displacement from the equilibrium position
of the $j$th particle, respectively, and $\lambda$ is a positive constant. In these two models of certain \textit{finite size}, the nontrivial resonances were found to be six-wave interactions, which lead to $T^{eq}\propto\lambda^{-8}\epsilon^{-4}$ \cite{Onorato4208} and $T^{eq}\propto\lambda^{-4}\epsilon^{-4}$ \cite{PhysRevLett.120.144301}, respectively, where $\epsilon$ is the energy density.

However, the equipartition hypothesis in its original form, i.e., for general systems in the thermodynamic limit, has not been proved. To prove the hypothesis in its original form is fundamentally important, because statistical mechanics as well as solid state theories are established in the thermodynamic limit. This issue has been mentioned in Refs \cite{Onorato4208,PhysRevLett.120.144301}, where the authors have conjectured that the four-wave resonances should dominate the irreversible dynamics in the thermodynamic limit for the FPU-$\alpha$ model and FPU-$\beta$ models, which result in $T^{eq}\propto\lambda^{-4}\epsilon^{-2}$ \cite{Onorato4208} and $T^{eq}\propto\lambda^{-2}\epsilon^{-2}$ \cite{PhysRevLett.120.144301}, respectively. These conjectures have not been verified. More importantly, the extension to general models is necessary. Another fundamental question is \textit{whether there is a universal scaling law} for the equipartition time in general 1D lattices. Various scaling laws of equipartition time have been reported previously \cite{PhysRevE.51.2877,BERCHIALLA2004167}. Nevertheless, they disagree with each other even for a specific model, e.g., FPU-$\alpha$-$\beta$ model \cite{Benettin2011,Benettin2013}.

In this Letter, based on wave resonance theory, we first show that there is a universal scaling law of the equipartition time for models given by Eq. (\ref{eqH}) \textit{in the thermodynamic limit}, i.e., $T^{eq}\propto\lambda^{-2}\epsilon^{-(n-2)}$ for $n\geq4$. Our key finding is that the lowest-order resonances,  i.e., the $n$-wave resonances for the model with a power law potential of power $n$, can take place in the thermodynamic limit, except when $n=3$. Our extensive numerical simulations confirm that this scaling is accurate for even $n$, but is slightly lower for odd $n$. There is a slight deviation but it decreases with the increase in $n$. This deviation suggests that there is an additional mechanism in the models with asymmetric interactions. By analyzing the deviation carefully and thoroughly, the existence of this additional mechanism is confirmed and is attributed to the Chirikov overlap \cite{1966SPhD1130I,CHIRIKOV1979263}. This is an astonishing new finding which implies that the Chirikov overlap may play a role in the weakly nonlinear regime. Finally, a universal scaling of $T^{eq}$ for a general interaction potential is derived and numerically verified.

{\it Theoretical analysis.}---We consider a lattice of $N+1$ particles with fixed ends ($q_0=q_{N}=0$) such that there are $N-1$ moving particles in between. The displacement of the $j$th particle can be written in terms of normal modes
\begin{equation}
  q_j=i\sum_{k=-N}^N\dfrac{Q_k}{\omega_k}e^{-ijk\pi/N},
\end{equation}
where $\omega_k=2|\sin(\frac{k\pi}{2N})|$ is the dispersion relation, and $Q_k$ is the amplitude of the $k$th normal mode \cite{BIVINS197365}. The boundary conditions, along with the reality of $q_j$, i.e., $q_j=q_j^{*}$, impose the constraint to the modes that $Q_k=Q_{-k}=Q_k^{*}$, and $Q_N=Q_{-N}=Q_0=0$. It is convenient to introduce the dimensionless complex amplitude of the $k$th normal mode
\begin{equation}\label{eqAk}
  a_k=\frac{\sqrt{N}Q_k+i\omega_k P_k/\sqrt{N}}{\epsilon^{1/2}\sqrt{2\omega_k}},
\end{equation}
where $P_k=\partial H/\partial\dot{Q}_k$ is the canonically conjugate momentum. Then, the Hamiltonian (\ref{eqH}) can be rewritten in the dimensionless form:
\begin{align}
  &\tilde{H}=H/\epsilon=\notag\\
  &\sum\omega_k a_ka_k^{*}+\frac{\lambda\epsilon^{\frac{n-2}{2}}}{n}\sum\Phi_{k_1}^{k_{n}}\delta(k_{1,n})\prod_{l=1}^n(a_{k_l}+a_{k_l}^{*}),
\end{align}
where $\Phi_{k_1}^{k_{n}}=\frac{N}{(2N)^{n/2}}
\frac{\sqrt{\prod_{l=1}^n\omega_{k_l}}}{\mathrm{sign}\left(\prod_{l=1}^nk_l\right)}$ is an interaction tensor coefficient, and $\delta(k_{1,n})$ gives the $n$-wave resonant condition for the wave vectors \cite{2011LNP825N}, i.e., $k_1\pm k_2\pm\cdots\pm k_n=0$. Whether the function sign takes $+1$ or $-1$ depends on the type of the $n$-wave process. Then, the equation of motion for the $k_1$th complex normal mode reduces to
\begin{equation}\label{eqWW}
    i\frac{\partial a_{k_1}}{\partial t}=\omega_{k_1}a_{k_1}
    +\lambda\epsilon^{\frac{n-2}{2}}\sum\Phi_{k_1}^{k_{n}}\delta(k_{1,n})\prod_{l=2}^n(a_{k_l}+a_{k_l}^{*}).
\end{equation}
From this equation we see that the nonlinear interactions are manifested as $n$-wave scattering terms. To evaluate the equipartition time, we introduce the wave action spectral density $A_i\delta_i^j=\langle a_{k_i}a^{*}_{k_j}\rangle$ following the wave resonance approach \cite{Onorato4208,PhysRevLett.120.144301}, where the brackets indicate the ensemble average and $\delta_i^j$ is the Kronecker delta. Based on the wave resonance theory \cite{2011LNP825N}, one can derive the $n$-wave kinetic equation
\begin{align}\label{eqAA}
  \dot{A}_1=4\pi\lambda^2\epsilon^{n-2}\int_{-\pi}^{\pi}|\Phi_{k_1}^{k_{n}}|^2\mathscr{F}(A_{1,n})\delta(k_{1,n}) \delta(\omega_{1,n}) dk_2\cdots dk_n,
\end{align}
where $\mathscr{F}(A_{1,n})$ is a function of $A_1, A_2, \cdots, A_n$, and $\delta(\omega_{1,n})$ gives the $n$-wave resonant condition for the frequencies, i.e., $ \omega_{k_1}\pm \omega_{k_2}\pm\cdots\pm\omega_{k_n}=0$ (see Sec. A of the Supplemental
Material (SM) Ref. \cite{SM} and Ref. \cite{2011LNP825N} for details). The summation of the wave vector from $-N$ to $N$ is replaced by an integral from $-\pi$ to $\pi$ on the reduced wave vector because the wave numbers in the Fourier space become dense and continuous in the thermodynamic limit.

Based on this evolution equation, $T^{eq}\propto\lambda^{-2}\epsilon^{-(n-2)}$ only holds when the nontrivial $n$-wave resonances exist and dominate the thermalization \cite{2011LNP825N,Onorato4208}. In the thermodynamic limit, the Fourier space of wave vectors becomes dense and the resonant conditions are not forbidden by the dispersion relation for $n\geq4$; therefore, resonant solutions must exist. Besides, considering the fact that any frequency has a certain broadening due to the nonlinearity, the resonant $n$-tuplets are interconnected.

For $n=3$, i.e., the FPU-$\alpha$ model, the three-wave resonances are forbidden because of the shape of the dispersion relation \cite{Onorato4208,Majda1997}. Hence, for this model one has to introduce a new canonical transformation to consider higher order interactions. We agree with the argument in Ref. \cite{Onorato4208} that the four-wave resonances dominate the thermalization in the thermodynamic limit, which leads to $T^{eq}\propto\lambda^{-4}\epsilon^{-2}$ .

Such an approach can be extended to more general symmetric potentials, $V(x)=|x|^d/d$, with $d=m_1/m_2>2$ and $m_1$ and $m_2$ are coprime. This is because the potential can be rewritten in terms of normal modes,
\begin{align}\label{eqVSym}
&\sum_{j}|q_j-q_{j-1}|^{m_1/m_2}=\notag\\
&\sum_{j}\left[\sum \frac{Q_{k_1}Q_{k_2}\cdots Q_{k_{2m_1}}}{\mathrm{sign}\left(k_1k_2\cdots k_{2m_1}\right)} e^{\frac{i\pi\left(\frac{1}{2}-j\right)\left(k_1+\cdots+k_{2m_1}\right)}{N}}\right]^{\frac{1}{2m_2}},
\end{align}
and the equation of motion can be obtained similarly. Equation (\ref{eqVSym}) indicates that the lowest number of waves participating in the scattering process is $2m_1$ in a model with exponent $d$, and $2m_1\geq4$ since $m_2\geq1$. Thus, the time scale for equipartition is $T^{eq}\propto\lambda^{-2}\epsilon^{-(d-2)}$ for such symmetric models only if the $2m_1$-wave resonances exist and dominate the irreversible dynamics in the thermodynamic limit.
\begin{figure*}[ht]
  \centering
  \includegraphics[width=1.6\columnwidth]{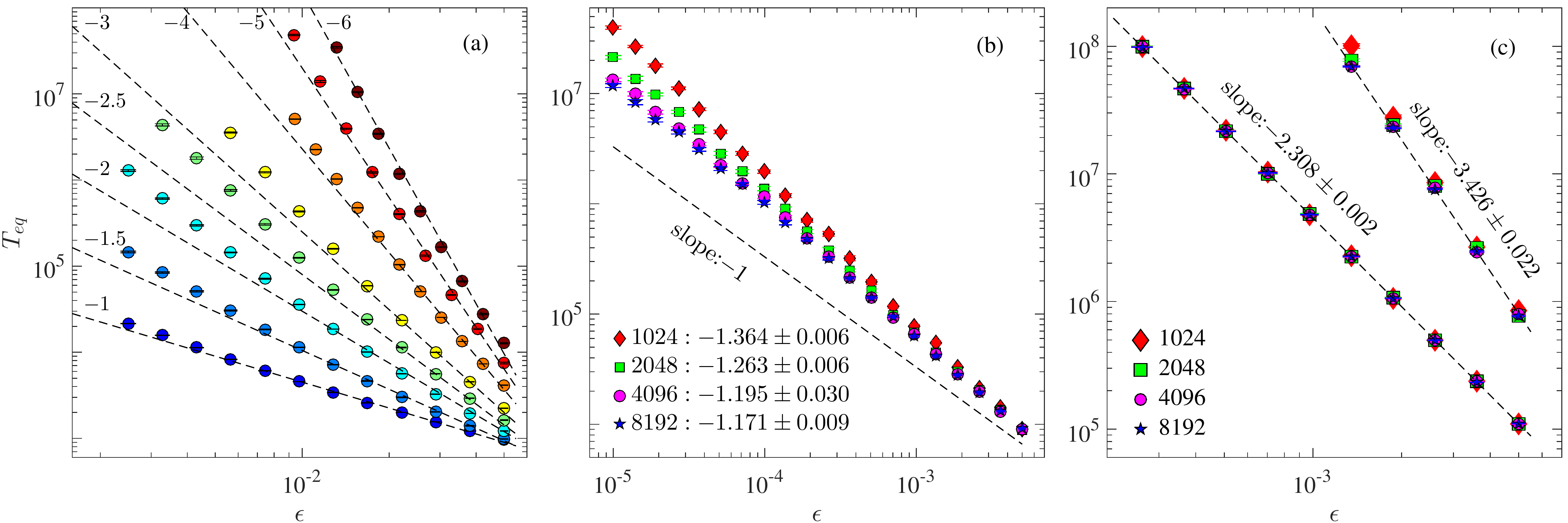}\\
  \caption{The equipartition time $T_{eq}$ as a function of energy density $\epsilon$ in log-log scale. (a) for symmetric models $V(x)=|x|^d/d$ with $d=3,7/2,4,9/2,5,6,7,8$ from bottom to top, and dashed lines with slope $2-d$ are drawn for reference, fixed $N=2048$. (b) for the symmetric model $d=3$ with different size $N=1024,2048,4096,8192$ from top to bottom. The slopes of best linear fit are listed in the plot, and a dashed line is drawn for reference. (c) for asymmetric models $n=3$ (bottom) and $n=5$ (top) with different size, and dashed lines are the best linear fit corresponding to $N=8192$. }\label{fig:Teq}
\end{figure*}

{\it Numerical experiments.}---Any numerical simulation is performed with finite size, and therefore, wave numbers are discrete in principle. However, the broadening effect of frequencies will lead to near-resonance interactions \cite{PhysRevLett.95.264302}. Therefore, one expects to approach the theoretical prediction in numerical simulations for lattices with an adequately large size. We adopt the method that is presented in Ref. \cite{Benettin2011} to calculate equipartition time. The normal modes of a 1D lattice of $N+1$ particles are: $Q_k=\sqrt{{2}/{N}}\sum_{j=1}^Nq_j\sin\left({jk\pi}/{N}\right)$, $P_k=\sqrt{{2}/{N}}\sum_{j=1}^Np_j\sin\left({jk\pi}/{N}\right)$. The energy of the $k$th normal mode is $E_k=\left(P_k^2+\omega_k^2Q_k^2\right)/2$. The indicator of thermalization, $\xi(t)=\tilde{\xi}(t)\frac{e^{\eta(t)}}{N/2}$, is adopted, where $\eta(t)=-\sum_{k=N/2}^{N}w_k(t)\log[w_k(t)]$ is the spectral entropy, in which $w_k(t)=\frac{\bar{E}_k(t)}{\sum_{l=N/2}^N\bar{E}_l(t)}$, $\tilde{\xi}(t)=\frac{\sum_{k=N/2}^N\bar{E}_k(t)}{\frac{1}{2}\sum_{1\leq{k}\leq{N}}\bar{E}_k(t)}$, and $\bar{E}_k(T)=\frac{1}{(1-\mu)T}\int_{\mu T}^TE_k(P(t),Q(t))dt$ is the average energy of the $k$th normal mode. Here, $\mu$ is a free parameter that controls the size of the time window for averaging. The equipartition time is measured as that satisfying $\xi(T_{eq})=1/2$.

For the numerical integration of Hamilton's canonical equations, we used the eighth-order Yoshida method \cite{YOSHIDA1990262}. To suppress fluctuations, the average is done over $60$ phases uniformly distributed in $[0,2\pi]$.

Note that $H'=\epsilon H$ under the scaling transformation $q'=q \epsilon^{1/2}$ for the power-potential models (\ref{eqH}); hence, the nonlinear parameter $\lambda$ and the energy density $\epsilon$ has a rigid scaling relation $\lambda'=\lambda \epsilon^{(n-2)/2}$. Therefore, it is equivalent to studying the scaling of $\lambda$ by fixing $\epsilon$ or that of $\epsilon$ by fixing $\lambda$. Here, we perform the latter with fixed $\lambda=1$. Figure \ref{fig:Teq}(a) shows the results for several symmetric power potentials with system size $N=2048$. It shows that the scaling $T^{eq}\propto \epsilon^{-(d-2)}$ agrees with the data very well, though a slight deviation can be recognized after a close look, which has been found to be a finite-size effect. This finite-size effect is shown in Fig.~\ref{fig:Teq}(b) by taking the case of $d=3$ as an example, where we can see that the larger the system size, the smaller the deviation; meanwhile, the lower the energy density, the larger the size must be to converge to the theoretical scaling.

Figure \ref{fig:Teq}(c) presents the results for models with $n=3$ and $n=5$. Best fitting gives $T^{eq}\propto\epsilon^{-2.31}$ and $T^{eq}\propto\epsilon^{-3.43}$, respectively. The finite-size effect is negligible comparing to that of symmetric potentials. The result for $n=3$ approaches the four-wave resonance prediction of $T^{eq}\propto\epsilon^{-2}$, while that for $n=5$ is close to the five-wave resonance prediction, i.e., $T^{eq}\propto\epsilon^{-3}$.

\begin{figure}[t]
  \centering
  \includegraphics[width=1\columnwidth]{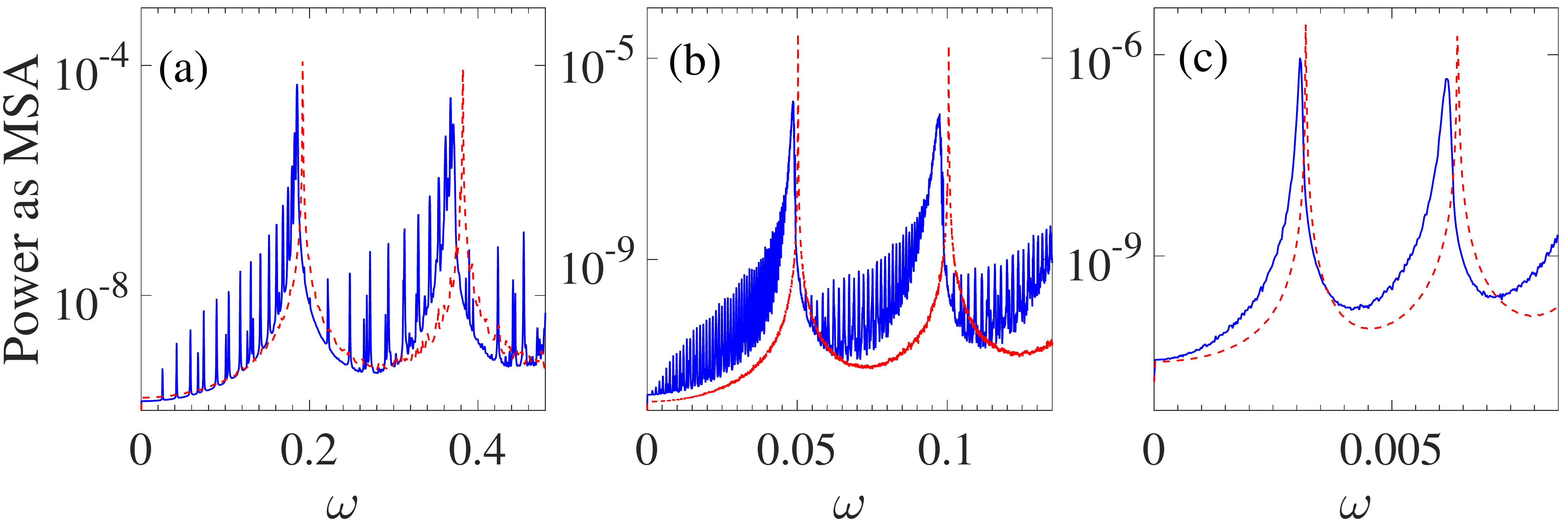}\\
  \caption{Power spectra of the momentum time series. Solid lines and dashed lines correspond to asymmetric ($n=3$) and symmetric ($d=3$) models, respectively. (a), (b) and (c) correspond to the lattice size $N=17$, $65$, and $1025$. $\epsilon=3\times10^{-3}$ is fixed.}\label{fig:FFT}
\end{figure}

To reveal why there is a deviation from the wave resonance theory for models with an odd exponent $n$, we study the power spectrum of a trajectory. Figures \ref{fig:FFT}(a)-(c) show the results for the FPU-$\alpha$ model with three sizes, $N=17$, $65$, and $1025$ respectively, at the fixed energy density $\epsilon=3\times10^{-3}$. For the sake of clarity, only the first two lowest frequency modes are drawn. Figures \ref{fig:FFT}(a) and \ref{fig:FFT}(b) show that there are many regularly distributed small peaks between two neighbouring normal modes in the asymmetric case ($n=3$), but they disappear in the symmetric case ($d=3$, see Sec. B of the SM \cite{SM} for details). Obviously, it is the symmetry of the inter-particle potential that makes the difference. By comparing the numbers of the small peaks for different system sizes, we find that the number of small peaks is $N-2$. This only depends on system size $N$ because the frequency difference between two neighboring normal modes decreases as $\sim N^{-1}$, the average frequency difference between two adjacent small peaks should be $\Delta \omega\propto N^{-2}$ for large $N$ (see Sec. C of the SM \cite{SM} for details). In addition, the amplitude of these small peaks sensitively depends on the energy density (it decreases with the energy density as a power law) and the order $n$ of the nonlinear term of the potential [the larger $n$, the lower the small peaks (see Sec. C of the SM \cite{SM})].

With the above analysis we can explain the simulation results. First, the normal modes can accurately represent the dynamics of the system with a symmetric potential in the thermodynamic limit since they are the unique energy carriers. Therefore, wave resonance theory works for such a system. Second, for a sufficiently small lattice with asymmetric power function potential, the small peaks are sparse and isolated. Despite their existence, they do not influence the irreversible dynamics; thus, the simulation results still agree with the resonant wave prediction for the FPU-$\alpha$ model with sufficiently small size \cite{Onorato4208}. Third, as the system size increases, the small peaks become closer to each other (as $\sim{N^{-2}}$). As a result, Chirikov overlap between two neighboring small peaks will occur because they must have a nonzero width due to nonlinearity. As such, for large $N$, the spectrum becomes continuous [see Fig.~\ref{fig:FFT}(c)]; it appears in effect as the envelope of the dense small peaks and the normal mode peaks. In this way, an additional transport channel opens. The Chirikov overlap mechanism works, and one should expect that it contributes to irreversible dynamics, which results in deviation from the wave resonance prediction. Finally, the amplitudes of small peaks are smaller at the fixed energy density for models with large power of $n$ (see Sec. C of the SM \cite{SM}). Therefore, as $n$ increases, the deviation from the wave resonance prediction decreases accordingly.

{\it General formula.}---A general interaction potential can be expanded as the Taylor series, i.e., $V(x)=\sum V^{(n)}(0)x^n/n!$. The corresponding equation of motion is similar to Eq. (\ref{eqWW}), with expanding series of $n$-wave interactions terms. In the thermodynamic limit, four-wave resonances excited by the term of $n=3$, and other $n$-wave resonances excited by other terms of $n>3$ exist simultaneously.

In the weakly nonlinear regime, these scattering processes can be considered to be independent. To integrate their contributions, inspired by Matthiessen's Rule \cite{Srivastava:215326}, it is reasonable to conjecture that the combined relaxation time $T_{eq}$ should follow the rule:
\begin{equation}\label{eq:Teq}
  \frac{1}{T^{eq}}=\frac{1}{T_3^{eq}}+\frac{1}{T_4^{eq}}+\cdots,
\end{equation}
where $T_n^{eq}$ represents the relaxation time contributed by the potential term with power exponent $n$.

\begin{figure}[t]
  \centering
  \includegraphics[width=1\columnwidth]{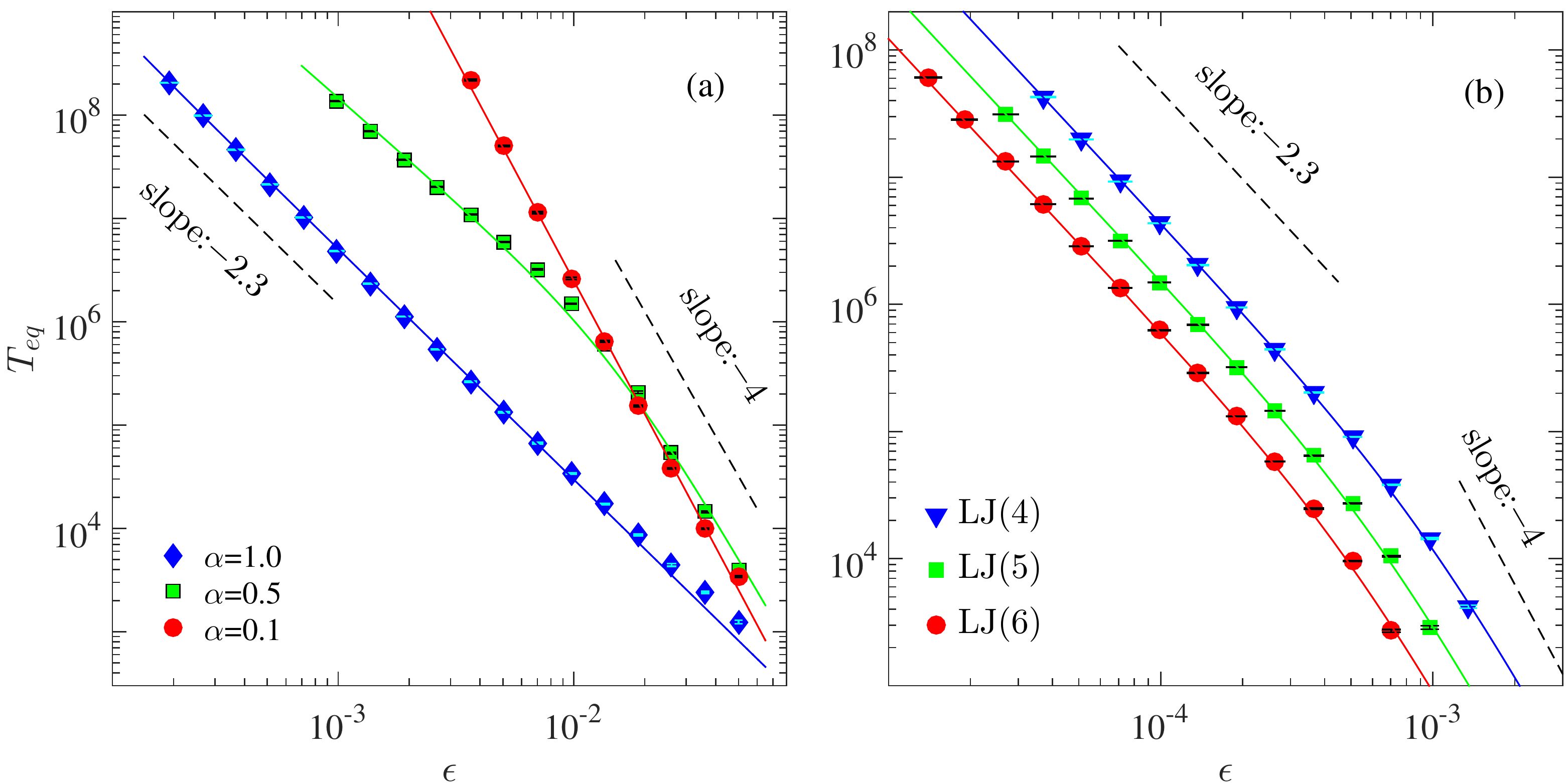}\\
  \caption{The equipartition time $T_{eq}$ as a function of energy density $\epsilon$ in log-log scale for (a) the cubic plus six-power potential model with different $\alpha$, and (b) the L-J model with $M=4,5,6$. Solid lines are best fitting with Eq. (\ref{eq:Teq}), and $n=3,6$ for (a); $n=3,4,5,6$ for (b). Dashed lines are drawn for reference.}\label{fig:TeqLJn}
\end{figure}

We check this conjecture with two numerical studies. The first one is for $V(x)=\frac{1}{2}x^2+\frac{\alpha}{3}x^3+\frac{1}{6}x^6$,
where $\alpha$ is used to adjust the relative weights of the two nonlinear terms. Here we introduce the sixth order nonlinearity instead of the fourth as the scaling exponents for $n=3$ and $n=4$ are too close ($T_3^{eq}\propto\epsilon^{-2.31}$ and $T_4^{eq}\propto\epsilon^{-2}$, respectively) to result in an obvious variation of the scaling exponent. Figure \ref{fig:TeqLJn}(a) shows simulation results for $\alpha=0.1$, $0.5$, and $1.0$, respectively. We see that the four-wave and six-wave resonance dominate the equipartition process in the two extremes of $\alpha=1$ and $0.1$, respectively. For a moderate cubic potential there appears a crossover. In all three cases, Eq. (\ref{eq:Teq}) can well fit the numerical results when we input $T_3^{eq}=c_3\epsilon^{-2.31}$ and $T_6^{eq}=c_6\epsilon^{-4}$ with proper weight parameters $c_3$ and $c_6$.

Our second example is the Lennard-Jones model \cite{1742-5468-2016-3-033205} which is frequently adopted for modeling a real lattice system, with $V(x)=\frac{1}{2M^2}\left[\frac{1}{(1+x)^{2M}}-\frac{2}{(1+x)^{M}}+1\right]$, where $M$ is an integer parameter. The numerical results are presented in Fig. \ref{fig:TeqLJn}(b). It shows that Eq. (\ref{eq:Teq}) fits the numerical results well for all values of $M$ we have tried.

{\it Conclusion.}---In summary, in models with interaction potential $V(x)=x^2/2+\lambda x^n/n$ the $n$-wave nontrivial resonances dominate the irreversible dynamics in the thermodynamic limit and lead to the universal scaling law of $T^{eq}\propto\lambda^{-2}\epsilon^{-(n-2)}$ for $n\geq4$. This scaling law can be extended to general symmetric models with $V(x)=x^2/2+\lambda|x|^d/d$, where $d$ is rational and $d>2$. Only for $n=3$ does one need a further canonical transformation to find higher-order resonances. Extensive numerical simulations confirm that this scaling holds perfectly in the symmetric models. It holds approximately for the asymmetric power potentials, but the deviation is slight.

Our models cover the most general class of 1D systems since any interaction potential can be expanded in terms of power potentials. This class of systems conserve both energy and momentum. Moreover, based on our scaling law, and inspired by Matthiessen's Rule, we have derived a universal scaling of $T^{eq}$ for a general interaction potential. An important conclusion is that a general nonlinear 1D lattice can be thermalized for arbitrary small nonlinearity.

To determine the mechanism of deviation from the universal scaling law in asymmetric models we established an important finding: that a large number of extra vibration modes are excited by the asymmetry of the potential, and Chirikov overlap may take place between them in a large system, which leads to the deviation. Furthermore, the extra vibration modes make phonon peaks asymmetrically broadened [see Fig. \ref{fig:FFT} and Sec. C of the SM \cite{SM}). This finding provides a new explanation for the asymmetric line shape of phonon spectra that has been widely reported in various condensed matter studies \cite{PhysRev.165.917,PhysRevB.59.1645,doi:10.1063/1.372199,PhysRevLett.109.046801,2017NatSR743602G,niehues2018strain,ceballoschuc2018influence,PhysRevB.97.195110}.  It is possible that more than one mechanisms are responsible for this phenomenon \cite{PhysRevLett.109.046801,2017NatSR743602G,niehues2018strain,ceballoschuc2018influence,PhysRevB.97.195110}.  That due to asymmetric interactions observed here should be more fundamental, as asymmetric interactions are general and ubiquitous in reality.

\begin{acknowledgments}
We are grateful to Jiao Wang for fruitful discussions. We acknowledge support by NSFC (Grant No. 11335006).
\end{acknowledgments}

\bibliography{final}

\end{document}